\documentclass[floatfix,aps,prd,twocolumn,showpacs,10pt]{revtex4}
\usepackage{epsfig}
\usepackage{epsf}
\usepackage{amsmath}
\usepackage{amssymb}
\usepackage{graphicx}

\begin{document}



\title{Cosmological Constraints on Radion Evolution  \\  in the Universal Extra Dimension Model}

\author{K. C. Chan$^{1,2}$ } \email{kcc274@nyu.edu}
\author{M.-C. Chu$^1$ } \email{mcchu@phy.cuhk.edu.hk}
\affiliation{$^1$   Department of Physics and Institute of Theoretical Physics, the Chinese University of Hong Kong, Hong Kong SAR, People's Republic of China 
\\
$^2$ Physics Department, New York University, NY 10003, USA  }

\date{\today}

\begin{abstract}

The constraints on the radion evolution in the Universal Extra Dimension (UED) model from Cosmic Microwave Background (CMB) and Type Ia supernovae (SNe Ia) data are studied. In the UED model, where both the gravity and standard model fields can propagate in the extra dimensions, the evolution of the extra dimensional volume, the radion, induces variation of fundamental constants. We discuss the effects of variation of the relevant constants in the context of UED for CMB power spectrum and SNe Ia data.  We then use the three-year WMAP data to constrain the radion evolution at $z \sim 1100$, and the 2 $\sigma$ constraint on $\dot{\rho} / \rho_0 $ ($\rho$ is a function of the radion, to be defined in the text) is $[ -8.8, 6.6] \times 10 ^{-13} $ yr$^{-1}$. The SNe Ia gold sample yields a constraint on $\dot{\rho} / \rho_0 $, for redshift between 0 and 1, to be $[-4.7, 14]\times 10^{-13} $ yr$^{-1}$. Furthermore, the constraints from SNe Ia can be interpreted as bounds on the evolution QCD scale parameter, $  $ $\dot{\Lambda}_{\rm QCD } / \Lambda_{\rm QCD, 0 }  $, $[-1.4, 2.8] \times  10^{-11} $ yr$^{-1}$, without reference to the UED model.

\end{abstract}


\maketitle

\section{Introduction}
\label{sect:Intro}

  In recent years, the interest in extra dimension models has been revived by string/M theory, in which the spacetime has to be multidimensional for the theory to be consistent \cite{Polchinski}. If large extra dimensions exist, the weakness of gravity may be ``explained'' by the large extra dimensional volume \cite{ADD}. Among the extra dimension models in the literature, the brane world (see \cite{Brane_review} for a review) model seems the most popular, where the standard model (SM) fields are trapped in a (3+1)-dimensional hypersurface, whereas gravity can freely propagate in the bulk.  Yet another class of extra dimension models where all the SM fields can propagate in the extra simensions has also been seriously considered. In this type of model, so called Universal Extra Dimension (UED) model, each SM particle is associated with a tower of Kaluza-Klein (KK) particles, with the mass difference between states equal to the inverse of the size of the extra dimensions.
This type of model is phenomenologically interesting because it can be investigated in  present and future collider experiments \cite{ACD}. Furthermore, the UED may explain the proton stability \cite{ProtonS}, the number of generations \cite{Generations}, neutrino masses \cite{NeutrinoM1, NeutrinoM2} and fermion chirality \cite{Chiral}. A cosmologically attractive feature of UED is that its lightest KK partner is stable and so it is a natural candidate for dark matter \cite{ServantTait,KongMatchev}.

      It is quite unnatural to think that the extra dimensional volume, called the radion, remains fixed per se, and so some mechanism must be initiated to stabilize the radion \cite{Chacko,Bucci}. However, in \cite{Bringmann}, it was shown that a broad range of standard stabilization mechanisms fail to admit a static solution in the UED model in the matter dominated regime. Besides, the redshifting of matter can also serve as a source to drive the evolution of the radion \cite{Perivolaropoulos}. Thus, we may envision that the radion does not sit at the minimum of the potential quietly at the epochs of CMB formation and SNe Ia explosions. In the UED scenario, the evolution of the radion will naturally induce the variation of various fundamental constants. We shall see that, in the Einstein frame, the fermion masses, the gauge conplings and the QCD scale parameter $\Lambda_{\rm QCD}$  will acquire radion dependence. Effects of variations of extra dimensional volume on the fundamental constants have also been discussed in Ref.~\cite{TDent,Correia}.

 By constraining the variation of constants at different redshifts, we can map out the allowed ranges of radion evolution at different epochs of the universe.  Constraint on radion evolution in the UED scenario by Big Bang Nucleosynthesis (BBN) has been obtained in \cite{BaojiuUED}.  In this paper, we seek complementary constraints from the Cosmic Microwave Background (CMB) and Type Ia supernovae (SNe Ia) data.

     Since the CMB anisotropies are measured quite accurately by WMAP \cite{WMAP3} and are sensitively dependent on various constants, we shall use the three-year WMAP data to constrain the radion evolution in UED models. Indeed, variations of several different fundamental constants have been constrained using CMB in the literature. Among the fundamental constants, the variation of the fine structure constant $\alpha$ receives the most attention \cite{Hannestad,Kaplinghat,Landau,Rocha}. Other constants constrained by CMB include the electron mass $m_e$ \cite{Yoo,Chan_me} and the gravitational constant $G$ \cite{Zahn,Umezu,Chan_G}.  However, unification theories \cite{Calmet,Langacker} and  string theory \cite{Ichikawa} suggest that the fundamental constants vary in a correlated way and should be constrained simultaneously. For the UED model at the epoch of CMB formation, the relevant constants $\alpha$, $m_e$ and $\Lambda_{\rm QCD}$ all depend on the radion.

     In the literature \cite{Amendola,Garcia-Berro99, Gaztanaga, Garcia-Berro06}, SNe Ia data have been used to constrain the variation of the gravitational constant  because the underlying physics of SNe Ia is dictated by the Chandrasekhar mass $M_{\rm Ch} \propto G^{-3/2}m_p^{-2}$, where $m_p$ is the proton mass. A similar method can be applied to constraining radion evolution because $m_p$ depends on the $\Lambda_{\rm QCD}$ and hence the radion.

However, when the radion is light enough to evolve cosmologically, its mass should be comparable to the Hubble parameter $H_0 \sim 10^{-34} \textrm{eV}$. Furthermore, the coupling of the radion to matter is not expected to be much weaker than that of gravity to the matter.  This means this field is highly constrained by tests of relativity and fifth force experiments \cite{Adelberger, Will}.   In particular, the solar system tests have ruled out such light scalar fields in the simplest models. A possible way out of this constraint is that the radion may be an environment-dependent chameleon field \cite{chameleon, Brax}. In this scenario, the effective potential of the scalar field depends on the density of the matter. On earth, the density is large and the the field is massive, and it can be as light as $H_0 $ in the cosmological scale. Because of the ``thin-shell effect'' of the chameleon field, the effective coupling is much reduced, and the solar system constraints can be easily satisfied. The current constraints on such a light scalar field can be satisfied if the field is lighter than $10^{-3}$ eV \cite{chameleon}.

     The rest of the paper is organized as follows. In Section~\ref{sec:DR}, we carry out the conventional dimensional reduction to get the radion dependence of various constants in the context of UED.  We then present the radion dependence of various constants relevant for CMB and SNe Ia, and their effects on the CMB power spectrum and SNe Ia in Section~\ref{sec:EffectCMBSNe}. The numerical constraints on the radion evolution and discussions on our constraints are presented in Section~\ref{sec:NConstraint}. We summarize in Section~\ref{sec:Conclusion}.

\section{Dimensional reduction and low energy effective actions}
\label{sec:DR}
  In this section we will start from the higher dimensional action and carry out the dimensional reduction to derive the dependence of the constants on the radion. The approach here follows closely that in Ref. \cite{Mazumdar,BaojiuUED}. Readers who are familiar with this type of set-up may go to Section~\ref{sec:EffectCMBSNe} directly. We shall dimensionally reduce the higher dimensional gravitational action to obtain its low energy effective action first. The higher dimensional action reads
\begin{equation}
\label{EqAction}
S=  \int d^{4+n}X \sqrt{-G}\left[  \left( \frac{1}{2 \kappa_{4+n}^2 } {}^{(4+n)}R\right) + \mathcal{L}_m \right],
\end{equation}
where $n$ denotes the number of compact extra dimensions, and $X^A$ represents the bulk spacetime coordinate, $A=0, 1, \dots, 3+n$. The higher dimensional Ricci scalar is denoted by ${}^{(4+n)}R$, $G$ is the determinant of the full spacetime metric $G_{AB}$, and $\kappa_{4+n}^2$ is a constant related to the higher dimensional Planck mass. $\mathcal{L}_m$ is the matter field Lagrangian density, which may include scalar fields, gauge fields and Dirac fermion fields.

     To proceed we shall take the metric ansatz
\begin{equation}
\label{EqMetric}
ds^2= G_{AB}dX^A dX^B= g_{\mu \nu}(x)dx^{\mu} dx^{\nu} + h_{i j}(x)dy^i dy^j.
\end{equation}    
In this metric the Greek indices run over 0, 1, 2 and 3, while the Latin indices $i$ and  $j$ run over the universal extra dimensions from 4 to $3+n$. We are only interested in the zero-mode of the KK expansion, and so the extra dimensional metric $h_{ij}(x)$ does not depend on the extra dimension coordinates. Furthermore, we have assumed that the metric in  Eq.~\ref{EqMetric} is block-diagonal because the vector-like connection $G^{i}_{\mu}$ vanishes for zero-mode.  The extra dimensions are compactified on an orbifold and the dimensionless coordinate $y^i$ assumes values in the interval [0,1]. 
     
 After expressing  ${}^{(4+n)}R$ in terms of the 4D Ricci scalar ${}^{(4)} R$ and integration by parts once, we obtain
\begin{eqnarray}
 S &  = &  \int d^{4}x d^n y \sqrt{-g}  \sqrt{h} \left[  \frac{1}{2 \kappa_{4}^2 \mathcal{V}_0 }   \left( {}^{(4)}R + \frac{1}{4}g^{\rho \sigma} \partial_{\rho} h^{i j} \partial_{\sigma} h_{i j}   \right.  \right.   
\nonumber \\
   &  +  &     \left.  \left.    \frac{1}{4}g^{\rho \sigma} h^{i j} \partial_{\rho}h_{i j} h^{k l} \partial_{\sigma}h_{k l} 
 \right) + \mathcal{L}_m \right],
\end{eqnarray}
where we have defined 
\begin{equation}
 \frac{1}{ \kappa_{4}^2} \equiv  \frac{ \mathcal{V}_{0} }{ \kappa_{4+n}^2 },
\end{equation}
with $\mathcal{V}_0$ being the volume of the extra dimensions today.

   We shall work in the Einstein frame with pure Ricci scalar in the gravitational action. To this end we will apply the conformal transformation
\begin{equation}
\label{Eq:Conformal}
\tilde{g}_{\mu \nu} =e^{-2 \theta } g_{\mu \nu}
\end{equation}
with 
\begin{equation}
\label{eq:ConformalRelation}
e^{-2 \theta} = \frac{\sqrt{h}}{\mathcal{V}_0}.
\end{equation}
After some algebra, we obtain the effective action
\begin{eqnarray}
\label{Eq:EffGrav}
 S    &   =  &    \int d^{4}x  \sqrt{-\tilde{g}}  \left[  \frac{1}{2 \kappa_{4}^2 }\left( {}^{(4)}R + \frac{1}{4} \tilde{g}^{\rho \sigma}  \partial_{\rho} h^{i j} \partial_{\sigma} h_{i j}  \right.  \right.   
\nonumber \\
  &  -  &  \left.  \left. \frac{1}{8}\tilde{g}^{\rho \sigma} h^{i j} \partial_{\rho} h_{i j} h^{k l}\partial_{\sigma} h_{k l} \right) + e^{4 \theta}\mathcal{L}_m \right]. 
\end{eqnarray}

     For concreteness, we further assume that the extra dimensional manifold  is homogeneous and isotropic. Hence the extra dimension metric takes the simple form:
\begin{equation}
\label{Eq:EDMetric}
\mathrm{diag}(b^2,b^2,\ldots, b^2).
\end{equation}
 With the ansatz Eq.~\ref{Eq:EDMetric}, the effective action reads
\begin{equation}
S= \int d^{4}x  \sqrt{-\tilde{g}}  \left[  \frac{1}{2 \kappa_{4}^2 }\left( {}^{(4)}R -   \frac{1}{2} \tilde{g}^{\mu \nu}  \partial_{\mu} \sigma  \partial_{\nu}\sigma  \right)   + e^{4 \theta}  \mathcal{L}_m \right],
\end{equation} 
where we have defined a new scalar field, the radion $\sigma$, as
\begin{equation}
\label{eq:sigma_radion}
\sigma \equiv \frac{1}{\kappa_4} \sqrt{\frac{n+2}{2n}} \ln \frac{b^n}{\mathcal{V}_0}. 
\end{equation}
We note that an effective potential has to be introduced to stabilize the radion. The effective potential may include contributions from the bosonic and fermionic sectors in addition to the higher dimensional potential \cite{Bucci} or the Casimir energy \cite{Chacko}.

   We now study the matter sector in more details. 
Let us begin with the minimally coupled scalar field 
\begin{equation}    
S_{\rm Scalar} = \int d^4 x d^n y   \sqrt{-G}\left(   - \frac{1}{2}G^{A B} \partial_{A}\phi \partial_{B}\phi - V(\phi) \right) .
\end{equation}
Because we are only interested in the low energy zero mode of the KK tower, the field $\phi$ is independent of the extra-dimensional coordinates, which can be integrated out.
After the conformal transformation Eq.~\ref{Eq:Conformal}, redefining the field as $\tilde{\phi} =\sqrt{ \mathcal{V}_0 } \phi $ and a change of variable $\tilde{V}(\tilde{\phi}) = \mathcal{V}_0 V(\phi)$,  we arrive at the action of the scalar field in the standard form
\begin{eqnarray}
S_{\rm Scalar} &  = & \int  d^4 x  \sqrt{ -\tilde{g} }   \left[  -\frac{1}{2} \tilde{g}^{\mu \nu} \partial_{\mu } \tilde{\phi} \partial_{\nu}\tilde{\phi}           \right.  \nonumber   \\ 
 &  -   &     \left.           \exp \left( -\kappa_4 \sigma \sqrt{\frac{2n}{n+2} }  \right)   \tilde{V}(\tilde{\phi})    \right].
\end{eqnarray}
If the scalar field is taken to be the Higgs field $H$, because the multiplicative factor $\exp \left( -\kappa_4 \sigma \sqrt{\frac{2n}{n+2} }  \right)$ does not change the minimum of $\tilde{V}$, the Higgs vacuum expectation value, $\langle H \rangle$, is unchanged.

       For the gauge field, the action is given by
\begin{equation}
S_{\rm Gauge} = - \frac{1}{4g_{*}^2} \int d^4 x d^n y  \sqrt{-g} \sqrt{h}  G^{AB}G^{CD}   F_{AC} F_{BD}, 
\end{equation}
where $F_{AB}$ is the gauge-invariant field strength tensor in the bulk, and $g_{*}$ is the $4+n$ dimensional gauge coupling. Again we are only limited to the zeroth KK mode, which is the 4D effective field tensor $F_{\mu \nu}$. 
Carrying out the transformation Eq.~\ref{Eq:Conformal}, we have
\begin{eqnarray}
S_{\rm Gauge} & = &    - \frac{1}{4g_{*}^2} \int d^4 x  \sqrt{- \tilde{g} } \exp\left( \kappa_4 \sigma  \sqrt{ \frac{2n}{n+2}} \right)       \nonumber    \\    & \times &    \tilde{g}^{\mu \nu} \tilde{g}^{\rho \sigma}  \tilde{F}_{\mu \rho}  \tilde{F}_{\nu \sigma},
\end{eqnarray} 
where we have defined $\tilde{F}_{\mu \nu} = \mathcal{V}_0 F_{\mu \nu}$.  Note that the 4D effective gauge couplings $\tilde{g}_{*}^2$ 
\begin{equation}
\label{eq:Gauge_vary}
\tilde{g}_{*}^2  = g_{*}^2   \exp \left(- \kappa_4 \sigma  \sqrt{ \frac{2n}{n+2}} \right)
\end{equation}
is radion dependent.


Similar techniques can be applied to the zero mode of the Dirac field $\psi$ with mass $\hat{m}$:\begin{equation}    
S_{\rm Dirac}= \int d^4 x d^n y   \sqrt{-g} \sqrt{h}   \left( - i \bar{\psi}e^{i \mu} \Gamma_{i} D_{\mu} \psi - \hat{m} \bar{\psi}\psi \right) ,
\end{equation}
where $e^{i \mu}$ is the vielbein, $\Gamma_i$ is the gamma matrix in $4+n$ dimension, and $D_{\mu}$ is the covariant derivative. 
After the conformal transformation, we get the canonical action
\begin{eqnarray}
\label{FermionMass}
S_{\rm Dirac} & = & \int d^4 x \sqrt{-\tilde{g}}\left[- i \bar{\Psi}  \tilde{e}^{i \mu}   \Gamma_{i} \tilde{D}_{\mu} \Psi     \right.   
  \nonumber  \\   
 & - &       \left.     \hat{m} \exp\left(- \frac{\kappa_4}{2} \sigma  \sqrt{\frac{2n}{n+2}} \right) \bar{\Psi} \Psi   \right]
\end{eqnarray}
by redefining the field $\psi$ as
\begin{equation} 
\Psi = \exp \left( -\frac{\kappa_4}{4}\sigma  \sqrt{ \frac{2n}{n+2}} \right) \psi.
\end{equation}
We can read out the radion dependence of the fermion mass $m$ from Eq.~\ref{FermionMass}:
\begin{equation}
\label{eq:fermion_mass_radion}
m =  \hat{m}  \exp\left(- \frac{\kappa_4}{2} \sigma  \sqrt{\frac{2n}{n+2}} \right). 
\end{equation}
 Since the fermion mass is given by $y \langle H \rangle $ in the standard model, and that $ \langle H \rangle $ has no radion dependence,  we may conclude that $ y \propto    \exp \left( - \frac{\kappa_4}{2} \sigma  \sqrt{\frac{2n}{n+2}} \right)  $. 

In the above, we assume that electroweak symmetry is broken spontaneously by the condensation of the higher dimensional Higgs field ~\cite{SSB_Csaki}.  In \cite{Bucci}, the authors construct an SM-like model from a UED model with a Higgs field in 5D. In particular, to generate fermions with massive and chiral zero mode in 4D, they have to use two fermion fields with opposite chiralities in addition to a 5D scalar field. Yet the radion dependence of the fermions in their model is the same as that we obtain from the above simple analysis.  In the literature, there is another class of models, called Higgsless models, in which the zero modes of the KK tower of the fermions are removed by choosing suitable conditions at the extra dimensions rather than by the Higgs mechanism. For the Higgsless model considered in \cite{Bucci}, the radion dependence of the fermion mass is also the same as Eq.~\ref{eq:fermion_mass_radion}. However, we should mention that in other types of models, \textit{e.g.} the ``gauge-Higgs model'' in \cite{Scrucca}, the Higgs VEV may depend on the radion, and thus the radion dependence of the fermion may be different.


\section{Variation of constants in  UED models and its impacts on the CMB power spectrum and SNe Ia }
\label{sec:EffectCMBSNe}

     In the last section, we have formally derived the radion dependence of the gauge couplings and the fermion masses. Now we are in a position to discuss the radion dependence of the constants, with the understanding that we are interested in  the epoch of CMB formation and afterwards.  Here we shall first discuss the radion dependences of $\alpha$, $m_e$, and $\Lambda_{\rm QCD}$ in Sect.~\ref{Const_UED_Dependence}.  We then discuss in Sect.~\ref{Effect_CMB_SNe} the effects of variation of these constants on the CMB power spectrum and on SNe Ia respectively.

\subsection{Variation of the relevant constants in the UED models}
\label{Const_UED_Dependence}
 Recall that we work in the Einstein frame, and so $G$ is constant. For CMB, the relevant constants are the electron mass $m_e$, electronic charge $e$ and proton mass $m_p$, whereas $m_p$ is important for SNe Ia.

  Among the fermion masses affected by the variable radion, the variation of $m_e$ is most relevant during the epoch of CMB formation. Since the quark mass contribution to the proton mass (or neutron mass) is only a few per cents or so, we neglect the effect of variation of the radion on nucleon  masses. For notational convenience, we use the subscripts $_{\rm CMB}$ and $_{0}$ to denote the values at CMB recombination and today respectively, and we define 
\begin{equation}
\rho_{\rm CMB}^{-1/2} 
\equiv   \exp \left( - \frac{\kappa_4}{2} \sigma_{\rm CMB}   \sqrt{\frac{2n}{n+2}} \right), 
\end{equation}
Note that Eq.~\ref{eq:sigma_radion} implies $\sigma_0=0$ and so $\rho_0 =1  $, and thus we have
\begin{equation}
 m_{e, \rm CMB} = \rho_{\rm CMB}^{-1/2} m_{e, 0}.
\end{equation}

   From Eq.~\ref{eq:Gauge_vary}, we now derive the radion dependence of the fine structure constant $\alpha$ and the QCD scale parameter $\Lambda_{\rm QCD}$. In SM, it is conventional to define $\alpha_i = \tilde{g}_i^2 /4 \pi  $, $i$=1, 2, 3, to denote the U(1), SU(2), and SU(3) coupling constants respectively, and so we have $\alpha_i \propto  \rho^{-1}   $. Using the usual SM relation $ \alpha^{-1} =  \alpha_1^{-1} + \alpha_2^{-1}$, we deduce the radion dependence of $ \alpha_{\rm  CMB} $:
\begin{equation}
\alpha_{\rm  CMB} = \alpha_{0} \rho_{\rm CMB}^{-1}.
\end{equation}

   The variation of the strong coupling $\alpha_3$ induces variation of the QCD scale parameter $\Lambda_{\rm QCD}$. As in usual QCD, the running of $\alpha_3$ from some high energy scale $Q$ to some lower energy scale $\mu$  is given by the one-loop renormalization group equation
\begin{equation}
\alpha_3(Q^2) = \frac{ \alpha_3(\mu^2)}{1 + \frac{\alpha_3 (\mu^2)  }{12 \pi }   (33 - 2 n_f ) \ln \frac{ Q^2  }{ \mu^2  }   }, 
\end{equation} 
where $n_f$ is the number of quark flavors with mass less than $Q$. When $Q$ is sufficiently low, the effective coupling becomes very large, and this scale, $\Lambda_{\rm QCD}$, is defined as 
\begin{equation}
\label{eq:LQCD}
\Lambda_{\rm QCD} = \mu \exp \left[  \frac{-6 \pi }{ (33-2n_f ) \alpha_3(\mu^2 )    }\right].
\end{equation}
Making use of Eq.~\ref{eq:LQCD} and matching the coupling at the thresholds $m_c$ and $m_b$, $\Lambda_{\rm QCD}$ can be expressed in term of $\alpha_3(M_Z)$ at the predetermined energy scale of Z boson \cite{BaojiuUED,DentFairbairn}
\begin{equation}
\Lambda_{\rm QCD} = M_Z \left( \frac{m_b m_c}{M_Z^2} \right)^{2/27} \exp \left(  \frac{- 2 \pi}{9 \alpha_3(M_Z)}  \right). 
\end{equation} 
Thus we can relate $\Lambda_{\rm QCD,CMB}$ to its present value  $\Lambda_{\rm QCD,0}$:
\begin{equation}
\label{eq:Lambda_CMB_0}
\Lambda_{\rm QCD,CMB}=  \Lambda_{\rm QCD,0} \rho_{\rm CMB}^{-2/27} \exp \left[
\frac{ 2 \pi}{9 \alpha_{3,0} (M_Z) }(1  - \rho_{\rm CMB} ) \right].
\end{equation} 
Because the contribution from $\Lambda_{\rm QCD}$ dominates the nucleon mass, we simply take the nucleon mass to be proportional to $\Lambda_{\rm QCD} $. Variation of $m_p$ will affect both the CMB power spectrum and the SNe Ia.

\subsection{Effects of variation of the constants on the CMB power spectrum and SNe Ia  }
\label{Effect_CMB_SNe}
In this subsection, we study how the variation of the constants discussed in Sect.~\ref{Const_UED_Dependence} affect the CMB power spectrum and SNe Ia.

Since the effects of variation of $\alpha $ \cite{Hannestad, Kaplinghat,Landau, Rocha}, $m_e$ \cite{Yoo, Chan_me} individually, and $\alpha$ and $m_e$ together \cite{Ichikawa} have been discussed quite extensively in the literature, we shall briefly discuss their impacts only. The ionization history has been modeled accurately by the code RECFAST, in which the ionization fractions of hydrogen and helium, and the matter temperature $T_{M}$ are evolved (see \cite{RECFAST} and references therein). The variation of $\alpha$ and $m_e$ modify the ionization history by varying the binding energies of  hydrogen H$_{\rm I} $ and helium He$_{\rm II}$, Thomson cross-section $\sigma_{T}$, the recombination coefficients $\mathcal{R}$ and the two-photon decay rates of hydrogen and helium.  The binding energies scale as $\alpha^2 m_e$; $\sigma_{T}$ is proportional to $\alpha^2  m_e^{-2}$, and the two-photon decay rates vary as $\alpha^8  m_e$ \cite{BT}.  In the literature, the recombination coefficients for hydrogen and helium are usually parametrized as a function of matter temperature. The recombination coefficient  $\mathcal{R}$ can be expressed as \cite{Kaplinghat}
\begin{equation}
\mathcal{R}  = \sum_{n,l}^{*} 8 \pi (2l+1) \left(  \frac{k T_M }{2 \pi m_e} \right)^{\frac{3}{2}}  \exp \left( \frac{B_n}{kT_M}  \right) \int_{\frac{B_n}{kT_M}}^{\infty}  \frac{\sigma_{nl} y^2  dy }{e^y -1 },  
\end{equation}
where $B_n$ is the binding energy of the $n$th state and $\sigma_{nl}$ is the ionization cross-section. The asterisk indicates that the sum should be regulated. Using the fact that $B_n$ scales with $m_e \alpha^2$ and the ionization cross-section varies as $m_e^{-2} \alpha^{-1}$ \cite{Uzan}, we can derive \cite{Kaplinghat,Kujat}
\begin{eqnarray} 
\label{eq:R_me_alpha_Tm}
\frac{   \partial \mathcal{R} }{ \partial  \alpha } &  = &  \frac{2}{\alpha } \left( 2 \mathcal{R} -  T_M \frac{ \partial  \mathcal{R}  }{\partial T_M}  \right),   \\
\frac{   \partial \mathcal{R} }{ \partial m_e} &  = & - \frac{1}{m_e} \left( 2 \mathcal{R}  +  T_M \frac{\partial  \mathcal{R}  }{\partial T_M}  \right),
\end{eqnarray}
which enable us to get the radion dependence of $\mathcal{R}$ using the empirical fitting formula for $\mathcal{R}$.  
       
      Among these effects, the change in the binding energy of hydrogen, $B_n$, is the most significant, and so the qualitive changes in the power spectrum can be understood based on the change in $B_n$.

The variation of $\Lambda_{\rm QCD} $ causes the hadron masses to vary. This will cause the energy density of the universe to change, and hence the expansion rate. 
A consistent theory should address where the energy goes to or comes from. It may arise from interaction of the SM particles with the radion. But without being too model-dependent, we simply ignore the soure/sink of the baryon energy density due to the variation of $\Lambda_{\rm QCD}$.  A  marked effect of the variation of the baryon energy density is the change in the ratio of baryon to photon energy density, usually defined as
\begin{equation}
R_{\rm CMB} = \frac{ 3 \rho_{\rm b , CMB}  } {4  \rho_{\gamma,\rm  CMB }}, 
\end{equation} 
which is related to the sound speed $c_s$ as
\begin{equation}
c_s^2 = \frac{ 1} {3( 1 + R_{\rm CMB} ) }.
\end{equation}
The effects of baryon on the power spectrum through $R_{\rm CMB}$ are as follows \cite{HuDodelson}. The presence of baryons reduces $c_s$ and hence the sound horizon; more remarkably, it shifts the equilibrium point of oscillations, which breaks the symmetry between the odd and even peaks.

\begin{figure}
	\centering
	\includegraphics[angle =-90 , width=0.9\linewidth] {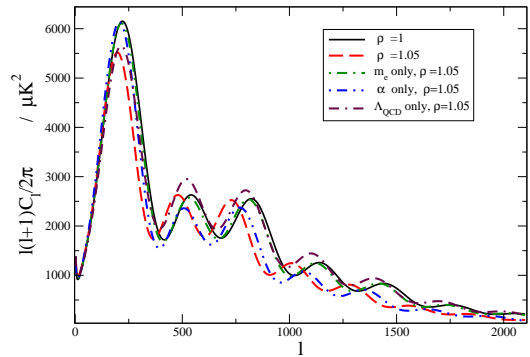}
        \caption{ The CMB temperature power spectrum with $\rho_{\rm CMB}=1.05$, and the individual contributions to the power spectrum due to the variations of $m_e$, $\alpha$ and $\Lambda_{\rm QCD}$ are plotted against the one with $\rho_{\rm CMB}=1$. See the text for discussions.}            
	\label{fig:UED_TT_com}
\end{figure}

 In Fig.~\ref{fig:UED_TT_com}, we show the power spectrum with $\rho_{\rm CMB}=1$ and $\rho_{\rm CMB}=1.05$. We also compare the effects of individual variations of $m_e$, $\alpha$ and $\Lambda_{\rm QCD}$ for $\rho_{\rm CMB}=1.05$ on the power spectrum in the figure. Note that for $\rho_{\rm CMB}=1.05$, $m_{e, \rm CMB}/m_{e,0}   $,   $\alpha_{\rm CMB}/ \alpha_{ 0}$ and $\Lambda_{\rm QCD, CMB}/ \Lambda_{\rm QCD, 0}$ are 0.98, 0.95 and 0.74 respectively. As the change in $m_e$ is small, we only see a small shift to lower $l$ scales due to later recombination caused by the smaller binding energy. Because the change in $\alpha$ and its weight on the binding energy of hydrogen are greater than those of $m_e$, the shifting of the peaks to small $l$ scales is more pronounced compared to changing $m_e$ alone. Later recombination implies that the visibility function is broadened, and so the recombination takes a longer time to finish and the damping effect becomes more significant in the large $l$ scales. We also expect the early integrated Sachs-Wolfe effect to be reduced for having less residual radiation during a later recombination.

 Because of the exponential factor in Eq.~\ref{eq:Lambda_CMB_0}, a large change in $\Lambda_{\rm QCD}$ results, which  means that the effect on the CMB power spectrum is appreciable. There are two major effects arising from the variation of $\Lambda_{\rm QCD}$.  First, a decrease in baryon density reduces the Hubble rate, and this implies that photon decoupling takes place later. The consequences of later recombination are that the sound horizon is larger, which corresponds to shifting of the peaks to smaller $l$ scales, and the damping in high $l$ scales is reduced. Second, a decrease in baryon mass causes the sound speed to increase, which not only results in a larger sound horizon, but also shifts the equilibrium point less. The latter helps to restore the symmetry between the odd and even peaks; in particular, the power of the first peak is reduced whereas the power of the second peak is boosted. Indeed, we see that peaks shift to smaller $l$ scales due to the larger sound horizon caused by these two effects.

The formation of CMB is basically determined by the values of $e$, $m_e$ and $m_p$ at the last scattering surface, and so the CMB constraint limits the variation of these constants during the CMB recombination epoch, $z \sim 1100$.

We now turn to discuss variation of the constants that are relevant for SNe Ia.  SNe Ia are believed to be explosions of white dwarfs that have approached the Chandrasekhar mass \cite{Hillebrandt}, $M_{\rm Ch} \propto G^{-3/2}m_p^{-2} $, which depends on the fundamental constants and is not sensitive to the structural details of the white dwarfs. Simple analytical formula for light curves of SNe Ia \cite{Arnett} predicts that the peak luminosity is proportional to the mass of nickel produced, which in turn is approximately proportional to $M_{\rm Ch}$.  In \cite{Gaztanaga}, the authors constrained the variation of the gravitational constant $G$ using SNe Ia data, and it was found that the effects of of varying $G$ on SNe Ia light curves are basically captured by the Chandrasekhar limit after comparing the results of simple $M_{\rm Ch}$ consideration  with full numerical computations. Similarly, varying the nucleon mass causes $M_{\rm Ch}$, and hence the intrinsic luminosity of the SNe Ia to evolve. Following \cite{Gaztanaga,Garcia-Berro06}, we shall constrain the variation of $\Lambda_{\rm QCD}$ using the $M_{\rm Ch}$ argument. Supposing that the luminosity $L \propto m_p^{-2}$, we have 
\begin{equation}
M = M_0 + 5 \log \frac{ m_p }{ m_{p,0 }  }, 
\end{equation} 
where $M$ is the absolute magnitude and the subscript $_0$ denotes the local value.

  Since there is no generic theory governing the evolution of the radion, in this paper, we shall parametrize the evolution of $\rho_{\rm SNe}$ as a linear function of redshift $z$:
\begin{equation}
\label{eq:Linrho}
\rho_{\rm SNe} = 1+ \zeta z.
\end{equation}  
Hence the apparent magnitude $m$ can be written as
\begin{equation}
\label{eq:ApparentMag}
m =   M_0 + 5 \log \frac{ m_p }{ m_{p,0 } } + 5 \log d_L +25,
\end{equation}
where $d_L $ is the luminosity distance, and $m_p \propto \Lambda_{\rm QCD, SNe}$ with $\Lambda_{\rm QCD, SNe}$ evaluated as in Eq.~\ref{eq:Lambda_CMB_0}.

\section{Numerical constraints on the evolution of the radion by CMB data and SNe Ia and Discussions }
\label{sec:NConstraint}  

In this section, we shall constrain the radion evolution using the three-year WMAP data \cite{WMAP3} and then the SNe Ia data \cite{Riess04}.

Since many free parameters are involved in CMB calculations, we make use of the Markov Chain Monte Carlo (MCMC) method implemented by the engine CosmoMC \cite{Cosmomc}, which searches for the region in the parameter space that maximizes the likelihood function, to derive the constraints on the free parameters. The theoretical CMB spectra are calculated by the Boltzmann code CMBFAST \cite{CMBFAST}. We vary the following set of parameters: the Hubble parameter, $ H_0$, the baryon density, $\omega_b = \Omega_b h^2$ ($h=H_0/100$ km s$^{-1}$ Mpc$^{-1}$), the cold dark matter density, $\omega_c =\Omega_c h^2 $, the reionization redshift, $z_{\rm re} $, the primordial fluctuation amplitude, $A_s$, the spectral index, $n_s$, and $\rho_{\rm CMB}$. Flatness of the universe is assumed in all the calculations.

\begin{figure}
	\centering
	\includegraphics[width=0.9\linewidth] {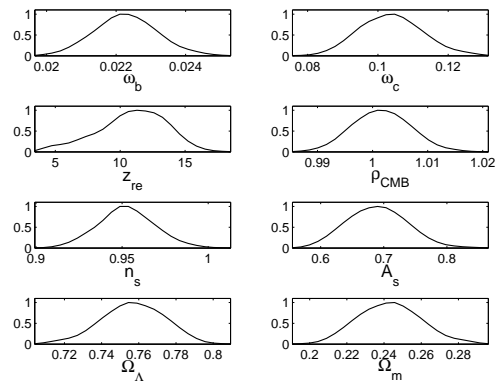}
        \caption{The marginal distributions of the free parameters, constrained by the three-year WMAP data. In addition to the free parameters $\omega_b \equiv \Omega_b h^2$, $\omega_c=\Omega_c h^2 $, $z_{\rm re}$, $n_s$, $A_s$ and  $\rho_{\rm CMB}$, shown also are the derived distributions of the density parameters of matter ($\Omega_m$) and cosmological constant ($\Omega_\Lambda$). The Hubble parameter is fixed to be 72 km s$^{-1}$ Mpc$^{-1}$. The maxima of the distributions are arbitrarily normalized to 1.}             
	\label{fig:UEDH072_dis}
\end{figure}

Assuming a top hat prior of [64,80] km s$^{-1}$ Mpc$^{-1}$ for $H_0$, we get the 95\% confidence interval (C. I.) for $\rho_{\rm CMB}$ to be [0.980, 1.021]. However, we find that $H_0$ is not constrained well within the assumed interval, and, more importantly, $H_0$ exhibits strong degeneracy with $\rho_{\rm CMB}$. To tighten the constraint on $\rho_{\rm CMB}$, we further fix $H_0$ to be the HST Key Project result, 72 km s$^{-1}$ Mpc$^{-1}$ \cite{HST} to break the degeneracy. The 95\% C.~I. for $\rho_{\rm CMB}$ now reduces to [0.991,1.012], and the marginal distributions of the parameters are shown in Fig.~\ref{fig:UEDH072_dis}, where we see that the standard cosmological parameters are constrained within the usual regions.

 In Ref.~\cite{Chan_me}, the radion evolution in the brane world scenario, in which only $m_e$ is expected to be radion dependent, is constrained by CMB data. Compared to that in the brane world model, the constraint on $\rho_{\rm CMB}$ in the UED scenario is tightened by about a factor of 2, thanks to the fact that $\alpha$ and $\Lambda_{\rm QCD}$, in addition to $m_e$, also vary.

     Since the upcoming Planck satellite mission is going to measure the temperature power spectrum to as high as $l \sim  2500$ and the $E$-polarization spectrum to $l \sim 1500$, we expect that there will be tremendous improvement in the constraint on $\rho_{\rm CMB}$. We can forecast the improvement that Planck will bring quantitatively using the Fisher matrix, which has been widely used to predict the expected uncertainties in future experiments (see \textit{e.g.} \cite{Eisenstein}). Under the assumption of Gaussian perturbations and Gaussian noise, the Fisher matrix takes the form
\begin{equation}
F_{i j} = \sum_{l} \sum_{X,Y} \frac{\partial C _{Xl} }{\partial p_i}  (\mathrm{Cov}_{lXY})^{-1} \frac{\partial C_{Yl} }{\partial p_j},
\end{equation}
where $p_i$ is the $i$th free parameter and $C_{X l}$ is the $l$th multipole of the observed spectrum of type $X$, which can be the temperature, temperature-polarization and $E$-polarization spectra. The experimental precision is encoded in the covariant matrix $\mathrm{Cov}_{lXY}$. Using the expected precision of the Planck satellite, we find that the current bound on $\rho_{\rm CMB}$ will be tightened by a factor of 8 or so when $H_0$ is fixed or free.

\begin{figure}
	\centering
	\includegraphics[width=0.9\linewidth] {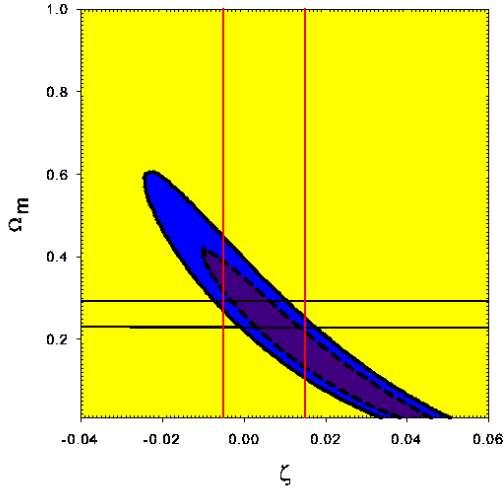}
        \caption{The joint $\chi^2  $ distribution of $\Omega_m$ and $\zeta$, constrained by SNe Ia data. The dashed and solid lines are the 68\% and 95\% condidence contours respectively. The horizontal (black) lines enclose the 1 $\sigma$ region of the Gaussian prior that we impose on $\Omega_m$, while the vertical (red) lines entail the 95 \% C.~I. on $\zeta$ after marginalizing over $\Omega_m$.      }             
	\label{fig:SNe_rho_2D}
\end{figure}

Now we turn to the constraint on the radion evolution by SNe Ia data. To be consistent with CMB calculations, we also assume flatness of the spatial curvature in evaluating $d_{L}$ in Eq.~\ref{eq:ApparentMag}; $H_0$ is analytically marginalized over \cite{Bridle}. 
 Assuming a linear variation of $\rho_{\rm SNe}$ as in Eq.~\ref{eq:Linrho}, we get the joint $\chi^2$ distribution of the density parameter of matter, $\Omega_m$, and $\zeta$, as shown in Fig.~\ref{fig:SNe_rho_2D}. Correspondingly, the 95\% C.~I. for $\zeta$ is [-0.025, 0.051].

   To tighten the constraint on $\zeta$, we impose a Gaussian prior on $\Omega_{m}$ using the WMAP plus SDSS constraint on $\Omega_{m}$, $0.266^{+ 0.026}_{-0.036}$ \cite{SpergelWMAP3}. We simply take the mean of the lower and upper error bars as 1 $\sigma$ for $\Omega_{m}$. The resultant 95\% C.~I. for $\zeta$, after numerically marginalizing over $\Omega_m$, is [-0.005, 0.015], or [0.998, 1.006] for $\rho_{\rm SNe}$.  The bounds on $\zeta$ are asymmetric due to the fact that the value of $\Omega_m$ derived from the SNe Ia gold sample \cite{Riess04} is higher than that favoured by WMAP and SDSS data (see \cite{SpergelWMAP3}).

    Because the constraint on $\rho_{\rm SNe} $ is solely due to the variation of $\Lambda_{\rm QCD}$, it is useful to translate the constraint on $\rho_{\rm SNe} $ to the constraint on  $\Lambda_{\rm QCD}$:
\begin{equation}
\label{eq:LQCDConstraint}
\Lambda_{\rm QCD, SNe} =\Lambda_{\rm QCD,0} (1 + \xi z),
\end{equation}  
with the 95\% C.~I. for $\xi$ to be [-0.30, 0.15]. We stress that Eq.~\ref{eq:LQCDConstraint} alone can be interpreted as a constraint on the variation of $\Lambda_{ \rm QCD}$, independent of UED model. 

    It is worth mentioning that we have also considered parametrizing the variation of $\Lambda_{\rm QCD}$ as Eq.~\ref{eq:LQCDConstraint} and confronting it with the SNe Ia data directly, and the resultant constraint on $\xi$ is similar to that derived from the constraint on $\zeta$. In other words, the bounds on $\Lambda_{\rm QCD} $ are not very sensitive to the two type of parametrizations, Eq.~\ref{eq:Linrho} and Eq.~\ref{eq:LQCDConstraint}.

Workers in SNe Ia community have been aware of the possibility of evolution of SNe Ia light curves. The light curves are usually adjusted based on the Phillips relation \cite{Phillips93} or other alternative schemes such as the stretch factor \cite{Perlmutter97}.  These schemes implicitly assume that there is one underlying parameter governing the light curves, although the nature of this parameter has not been clearly identified. However, scatter about the mean relation always exists, and a recent study exacerbates the inadequacy of the one-parameter description \cite{Benetti04}. Although most theorists suspect that another underlying parameter is related to the metallicity $Z$ of the progenitor \cite{DHS01, Timmes, Podsiadlowski}, observations so far do not reveal any definite $Z$ dependence \cite{Hamuy00, Gallagher}. So far most of the studies concentrate on metallicities or the two-component model \cite{Scannapieco,Howell}, we would like to emphasize that the second parameter may be due to variation of constants, such as $G$ \cite{Amendola, Garcia-Berro99, Gaztanaga} or $\Lambda_{\rm QCD}$.

More drastically, variation of constants may even mimick the effect of dark energy, but this is increasingly untenable as more and more evidences, such as the CMB, necessitate the existence of the cosmological constant or suchlikes.

   Our argument of the effect of variation of $\Lambda_{\rm QCD}$ on the light curves of SNe Ia rests on the fundamental Chandrasekhar mass; nonetheless it is just the first step and should be checked by detailed simulations that the fine details of the explosion per se do not give rise to significant corrections. This may be a daunting task given that numerous nuclear reaction cross-sections can be affected. See \cite{SNe_Fairbairn} for a discussion on the effect of varying strong coupling constant on the fusion reaction in the SNe Ia.

Finally, we would like to compare the constraints on $ \rho_z $ at different redshifts $z$. The constraints from BBN \cite{BaojiuUED}, CMB and SNe Ia are displayed in Table \ref{tab:rho_constraint_table}.   It is meaningful to consider $\dot{\rho}/\rho_0 =  ( 1  -   \rho_z  )/ |\Delta t| $.  For SNe Ia, we use the mean redshift of the gold sample, 0.4, to calculate $\Delta t $. For the constraints on $\Lambda_{\rm QCD}$ (Eq.~\ref{eq:LQCDConstraint}), they can be translated to the bounds on $\dot{\Lambda}_{\rm QCD } / \Lambda_{\rm QCD,0}  $: $[-1.4, 2.8] \times  10^{-11} $ yr$^{-1}$.  Again, we mention that the simplest light radion models have been ruled out by test of general relativity and fifth force experiments although we don't include them here.

\begin{table}[htb!]
\caption{  The constraints on $\rho$ from BBN, CMB and SNe Ia. } 
\label{tab:rho_constraint_table}
\begin{tabular}{|l |  l | l |  p{0.3\linewidth} | }
\hline
 Redshift $z$  &  Observations   & 95 \% C.~I. for $ \rho_z $   & 95 \% C. I. for   $ \dot{\rho} /  \rho_0  $  ($ 10^{-13}$    yr$^{-1}$)    \\

\hline
  $\sim 10^{10} $   &       BBN \cite{BaojiuUED}      &   [0.9992, 1.0027]   &  [-2.0, 0.59]   \\
\hline

$\sim  1100 $ &   CMB   &    [0.991, 1.012]  &  [-8.8, 6.6]    \\
\hline
$0 \sim 1 $  &   SNe Ia   &  [0.998, 1.006]  & [-14 , 4.7]      \\
\hline

\end{tabular}
\end{table}

\section{Conclusion}
\label{sec:Conclusion}  

In this paper, we consider variation of fundamental constants in the context of Universal Extra Dimension models, where not only the gravity, but also the standard model fields can propagate in the extra dimensions. We explicitly show the radion dependence of various constants by going through the dimensional reduction procedure. It turns out that the constants relevant for CMB power spectrum are $\alpha$, $m_e$ and $\Lambda_{\rm QCD}$, while it is $\Lambda_{\rm QCD}$ that is important for SNe Ia light curves. The effects of the variation of the constants on the CMB power spectrum and SNe Ia are discussed. For CMB, we have done detailed numerical calculations to get the power spectrum. The effect of variation of $\Lambda_{\rm QCD}$ on SNe Ia light curves is estimated based on the fundamental Chandrasekhar mass. We then use the three-year WMAP data to constrain the radion evolution at $z \sim 1100$, and the 2 $\sigma$ constraint on $\dot{\rho} / \rho_0 $ is $[ -8.8, 6.6] \times 10 ^{-13} $ yr$^{-1}$. The SNe Ia gold sample yields a constraint on $\dot{\rho} / \rho_0 $, at redshift from 0 to 1, to be $[-14 , 4.7]\times 10^{-13} $ yr$^{-1}$. Furthermore, the constraints from SNe Ia can be interpreted as bounds on $\dot{\Lambda}_{\rm QCD } / \Lambda_{\rm QCD,0}  $, $[-1.4, 2.8] \times  10^{-11} $ yr$^{-1}$.

\begin{acknowledgments}
We would like to thank T. Dent, M. Fairbairn, and T. Bringmann, and the anonymous referee for useful comments. We also thank the ITSC of the Chinese University of Hong Kong for providing its clusters for computations. This work is partially supported by a grant from the Research Grant Council  of the Hong Kong Special Administrative Region, China (Project No. 400707).    
\end{acknowledgments}



\begin{thebibliography}{99}
\small

        \bibitem{Polchinski}J. Polchinski, \textit{String Theory vol 1, 2} (Cambridge University Press, Cambridge, 2003). 

        \bibitem{ADD} N. Arkani-Hamed, S. Dimopoulos, and  G. Dvali, Phys. Lett. B \textbf{429}, 263 (1998); I. Antoniadis, N. Arkani-Hamed, S. Dimopoulos, and  G. Dvali, Phys. Lett. B \textbf{436}, 257 (1998); D. Cremades, L. E. Ibanez, and F. Marchesano, Nucl. Phys. B \textbf{643}, 93 (2002); C. Kokorelis, Nucl. Phys. B \textbf{677}, 115 (2004).

        \bibitem{Brane_review}C. Cs\'aki, arXiv:hep-ph/0404096; R. Maartens, arXiv:gr-qc/0312059 v2.
        \bibitem{ACD}T. Appelquist, H.-C. Cheng, and B. A. Dobrescu, Phys. Rev. D \textbf{64}, 035002 (2001).

        \bibitem{ProtonS}T. Appelquist,  B. A. Dobrescu, E. Ponton, and H. U. Yee, Phys. Rev. Lett. \textbf{87}, 181802 (2001).
        \bibitem{Generations} B. A. Dobrescu, and E. Poppitz, Phys. Rev. Lett. \textbf{87}, 031801 (2001).
            
        \bibitem{NeutrinoM1}T. Appelquist,  B. A. Dobrescu, E. Ponton, and H. U. Yee, Phys. Rev. D \textbf{65}, 105019 (2002).

        \bibitem{NeutrinoM2}R. N. Mohapatra, and A. Perez-Lorenzana, Phys. Rev. D \textbf{67}, 075015 (2003).

        \bibitem{Chiral}G. Burdman, B. A. Dobrescu, and E. Ponton, JHEP \textbf{0602}, 003 (2006).

         \bibitem{ServantTait} G. Servant, and T. M. P. Tait, Nucl. Phys. B \textbf{650}, 391 (2003).

         \bibitem{KongMatchev} K. Kong, and K. T. Matchev, JHEP \textbf{0601}, 038 (2006).

        \bibitem{Chacko}Z. Chacko, and P. Perazzi, Phys. Rev. D \textbf{68}, 115002 (2003).    
        \bibitem{Bucci}P. Bucci, B. Grzadkowski, Z. Lalak, and R. Matyszkiewicz, JHEP \textbf{04}, 067 (2004).

        \bibitem{Bringmann}T. Bringmann, and M. Eriksson, JCAP \textbf{10}, 006 (2003).    

        \bibitem{Perivolaropoulos}L. Perivolaropoulos, and C. Sourdis, Phys. Rev. D \textbf{66}, 084018 (2002).


        \bibitem{TDent}T. Dent, Nucl. Phys. B \textbf{677}, 471 (2004). 
        \bibitem{Correia}F. P. Correia, M. G. Schmidt, and Z. Tavartkiladze, arXiv:hep-ph/0211122v2.      




        \bibitem{BaojiuUED}B. Li, and M.-C. Chu, Phys. Rev. D \textbf{73}, 025004 (2006).   



	\bibitem{WMAP3} G. Hinshaw \textit{et al.}, arXiv:astro-ph/0603451; L. Page \textit{et al.}, arXiv:astro-ph/0603450. 




        \bibitem{Hannestad} S. Hannestad, Phys. Rev. D \textbf{60}, 023515 (1999).
        \bibitem{Kaplinghat}   M. Kaplinghat, R. J. Scherrer, and M. S. Turner, Phys. Rev. D \textbf{60}, 023516 (1999).

         \bibitem{Landau}  S. J. Landau, D. D. Harari, and M. Zaldarriaga, Phys. Rev. D \textbf{63}, 083505 (2001).
         \bibitem{Rocha} G. Rocha \textit{et al.}, MNRAS \textbf{352}, 20 (2004).


        \bibitem{Yoo} J. J. Yoo, and  R. J. Scherrer, Phys. Rev. D \textbf{67}, 043517 (2003).
        \bibitem{Chan_me}K. C. Chan, and M.-C. Chu, Phys. Rev. D \textbf{75}, 8 (2007).


        \bibitem{Zahn} O. Zahn, and M. Zaldarriaga, Phys. Rev. D \textbf{67}, 063002 (2003).
        \bibitem{Umezu} K. I. Umezu, K. Ichiki, and M. Yahiro, Phys. Rev. D \textbf{72}, 044010 (2005).
        \bibitem{Chan_G} K. C. Chan, and M.-C. Chu, Phys. Rev. D \textbf{75}, 083521 (2007).

       \bibitem{Calmet} X. Calmet, and H. Fritzsch, Eur. Phys. J. C \textbf{24}, 639 (2002).
       \bibitem{Langacker}P. Langacker, G. Segre, and  M. J. Strassler, Phys. Lett. B \textbf{528}, 121 (2002).



       \bibitem{Ichikawa}K. Ichikawa, T. Kanzaki, and M. Kawasaki, Phys. Rev. D \textbf{74}, 023515 (2006).


        \bibitem{Amendola}L. Amendola, S. Corasaniti, and F. Occhionero, arXiv:astro-ph/9907222.
        \bibitem{Garcia-Berro99} E. Garc\'ia-Berro, E. Gazta\~naga, J. Isern, O. Benvenuto, and L. Althaus, arXiv:astro-ph/9907440.

       \bibitem{Gaztanaga}E. Gaztanaga, E. Garc\'ia-Berro, J. Isern, E. Bravo, and I. Dom\'inguez, Phys. Rev. D \textbf{65}, 023506 (2001).

        \bibitem{Garcia-Berro06} E. Garc\'ia-Berro, K. Kubyshin, P. Lor\'en-Aguilar, and  J. Isern, Int. J. Mod. Phys. D \textbf{15}, 1163 (2006).





        \bibitem{Adelberger}E. G. Adelberger, B. R. Heckel, and A. E. Nelson, Annu. Rev. Nucl. Part. Sci. \textbf{53}, 77 (2003).
        \bibitem{Will}C. M.  Will, Living Rev. Rel.  \textbf{4}, 4 (2001). 

        \bibitem{chameleon}J. Khoury, and A. Weltman, Phys. Rev. Lett. \textbf{93}, 171104 (2004); Phys. Rev. D \textbf{69}, 044026 (2004).
        \bibitem{Brax} Ph. Brax, C. van de Bruck, A.-C. Davis, J. Khoury and A. Weltman, Phys. Rev. D \textbf{70}, 123518 (2004). 
  




        \bibitem{Mazumdar}A. Mazumdar, R. N. Mohapatra, and  A. P\'erez-Lorenzana, JCAP \textbf{0406}, 004 (2004).        

        \bibitem{SSB_Csaki}For a review of electroweak symmetry breaking in extra dimensions in the Higgs and Higgsless models see C. Cs\'aki, J. Hubisz, and P. Meade, hep-ph/0510275.

        \bibitem{Scrucca}C. A. Scrucca, M. Serone, and L. Silvestrini, Nucl. Phys. B \textbf{669}, 128 (2003).
  




        \bibitem{DentFairbairn}T. Dent, and M. Fairbairn, Nucl. Phys. B  \textbf{653}, 256 (2003).








        \bibitem{RECFAST}S. Seager, D. D. Sasselov, and  D. Scott, Astrophys. J. \textbf{523}, L1 (1999).        

        \bibitem{BT}G. Breit, and  E. Teller, Astrophys. J. \textbf{91}, 215 (1940).        
	\bibitem{Uzan}J. -P. Uzan, Rev. Mod. Phys. \textbf{75}, 403 (2003).


        \bibitem{Kujat} J. Kujat, and  R. J. Scherrer, Phys. Rev. D \textbf{62}, 023510 (2000).



         \bibitem{HuDodelson}W. Hu, and S. Dodelson, Annu. Rev. Astron. Astrophys. \textbf{40}, 171 (2002).










        \bibitem{Hillebrandt}W. Hillebrandt, and  J. C. Niemeyer,  Annu. Rev. Astron. Astrophys.  \textbf{38}, 191 (2000).

        \bibitem{Arnett}W. D. Arnett, Astrophys. J. \textbf{253}, 785 (1982).



        \bibitem{Riess04} A. G. Riess \textit{et al.}, Astrophys. J. \textbf{607}, 665 (2004).


        \bibitem{Cosmomc}A. Lewis, and  S. Bridle, Phys. Rev. D \textbf{66}, 103511 (2002).        
        \bibitem{CMBFAST}U. Seljak, and  M. Zaldarriaga, Astrophys. J. \textbf{469}, 437 (1996). 


        \bibitem{HST}W.L. Freedman \textit{et al.}, Astrophys. J. \textbf{553}, 47 (2001).        

	\bibitem{Eisenstein} D. J. Eisenstein, W. Hu \& M. Tegmark, Astrophys. J.  \textbf{518}, 2 (1999).  

        \bibitem{Bridle}S. L. Bridle \textit{et al.}, MNRAS \textbf{335}, 1193 (2002).        




        \bibitem{SpergelWMAP3} D. N. Spergel \textit{et al.}, arXiv:astro-ph/0603449 v2.


        \bibitem{Phillips93} M. M. Phillips, Astrophys. J. \textbf{413}, L105 (1993).        
        \bibitem{Perlmutter97} S. Perlmutter \textit{et al.}, Astrophys. J. \textbf{483}, 565 (1997).        

        \bibitem{Benetti04} S. Benetti \textit{et al.}, MNRAS \textbf{348}, 261 (2004).        



        \bibitem{DHS01}I. Dom\'inguez, P. H\"oflich, and O. Straniero, Astrophys. J. \textbf{557}, 279 (2001).        
        \bibitem{Timmes}  F. X. Timmes, E. F. Brown, and J. W. Truran, Astrophys. J. \textbf{590}, L83 (2003).        
   \bibitem{Podsiadlowski} P. Podsiadlowski \textit{et al.}, arXiv:astro-ph/0608324.        






        \bibitem{Hamuy00}M. Hamuy \textit{et al.}, Astronomical J. \textbf{120}, 1479 (2000).        
        \bibitem{Gallagher}J. S. Gallagher \textit{et al.}, Astrophys. J. \textbf{634}, 210 (2005).        


        \bibitem{Scannapieco}E. Scannapieco, and L. Bildsten, Astrophys. J. \textbf{629}, L85 (2005).       
        \bibitem{Howell} D. A. Howell, M. Sullivan, A. Conley, and R. Carlberg, arXiv:astro-ph/0701912.  


        \bibitem{SNe_Fairbairn}M. Fairbairn, arXiv:astro-ph/9910328.









    











            
%

































     



















       












        

















\end{thebibliography}
\end{document}